\documentclass{article}
\usepackage{arxiv}
\usepackage[utf8]{inputenc} 
\usepackage[T1]{fontenc}   
\usepackage{hyperref}      
\usepackage{url}            
\usepackage{booktabs}      
\usepackage{amsfonts}       
\usepackage{nicefrac}       
\usepackage{microtype}      
\usepackage{lipsum}
\usepackage{multicol}
\usepackage{graphicx}
\usepackage{graphics}
\usepackage{subfig}
\title{Quantifying Dismantlement in Disconnected Networks}

\author{
  Siddharth Patwardhan \\
  Department of Mathematics and Statistics\\
  Indian Institute of Science Education and Research Kolkata\\
  \texttt{sbp14ms080@iiserkol.ac.in} \\
 }
\begin{document}
\maketitle
\begin{abstract}
We propose a novel measure to quantify dismantlement of a fragmented network. The existing measure of dismantlement used to study problems like optimal percolation is usually the size of the largest component of the network. We modify the measure of uniformity used to prove the Szemeredi's Regularity Lemma to obtain the proposed measure. The proposed measure incorporates the notion that the measure of dismantlement increases as the number of disconnected components increase and decreases as the variance of sizes of these components increases.   
\end{abstract}
\section{Introduction}
Sometimes, a fragmented network is more desirable than a fully connected functioning one. Consider, for example, the need to eliminate bacteria by disrupting their molecular network or by vaccinating a few individuals in a population to break up the contact network through which an infection spreads\cite{DesPerf}. This enables us to ask the question, to what degree has a network been fragmented? It is known that the removal of a few highly connected nodes, or hubs, can break up a complex network into many disconnected components\cite{Err} and to find these nodes, is an important problem in network science. It also makes it necessary to quantify dismantlement to verify the effect these nodes have on the network upon removal, which is done by measuring the size of the largest component\cite{PercThre}\cite{InfMax}. It effectively quantifies dismantlement in the context of the optimal percolation of most real world networks. However, in case of the optimal percolation of many social, technological and transport networks which have a community structure\cite{sn}, we come across networks in which the largest component does not essentially quantify dismantlement. It can be seen from the networks illustrated in figure 1 that the size of the largest component does not always correctly quantify dismantlement in a disconnected network.

\begin{figure}[h]
    \centering
    \subfloat{\includegraphics[width=6.25cm]{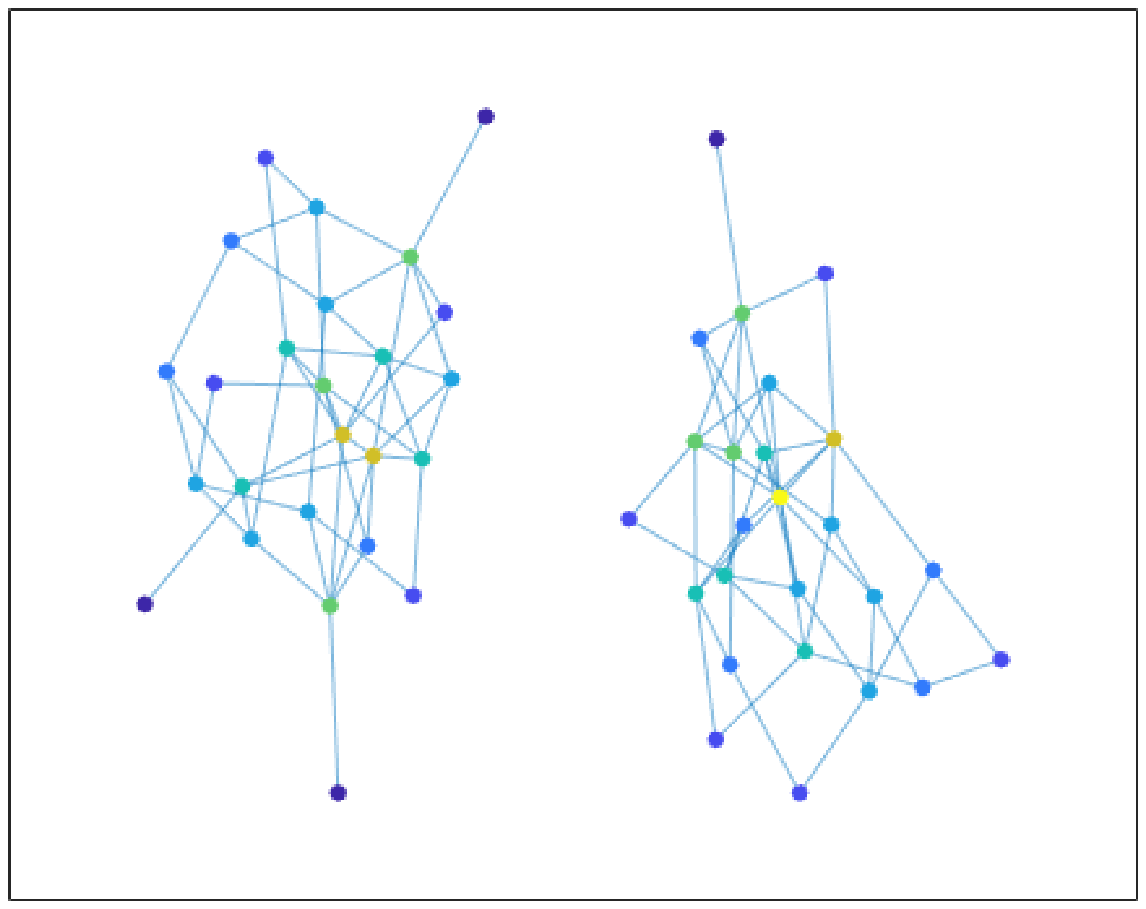}} 
    \qquad
    \subfloat{\includegraphics[width=6.25cm]{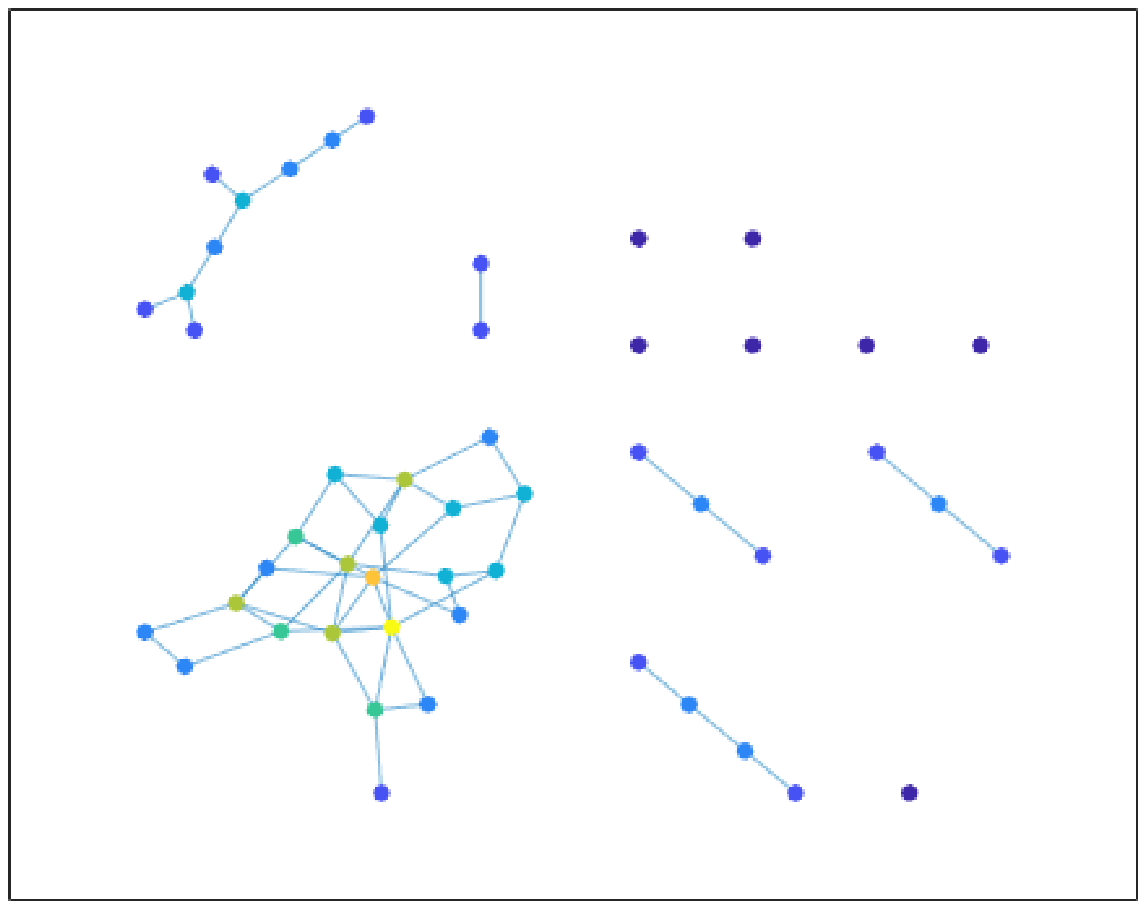}}
    \caption{Both the networks have a largest component of size 25, but the second network is clearly more dismantled than first network.}
    \label{fig:example}
\end{figure}

Almost 30 years ago, in the course of the proof of a major result on the Ramsey properties of arithmetic progressions, Szemeredi developed a graph theoretical tool, the regularity lemma, whose fundamental importance has been realized more and more in recent years. The lemma says that all graphs can be approximated by random graphs in the following sense: every graph can be partitioned, into a bounded number of equal parts, so that most of its edges run between different parts and the edges between any two parts are distributed fairly uniformly. Just as we would expect it if they had been generated at random\cite{gtbook}.

\section{Proposed Measure}
\label{S:2}
In the proof for Szemeredi's Regularity Lemma \cite{gtbook}, given a graph $G = (V,E)$ and $n=|V|$ with a partition $P = \{ V_0, V_1, .. ,V_k \}$, the elements in partition $P$ represent the sizes of sub-graphs. The uniformity of the partition is given by $$q(P) =\sum_{i<j} \frac{|V_i||V_j|d^2(V_i,V_j)}{n^2} $$ where, $d$ is the edge density between any two sub-graphs, given by $$ d(A,B)=\frac{||A,B||}{|A||B|}$$ where $||A,B||$ is the number of edges between the two sub-graphs. In the context of a disconnected graph, we consider the partition to be such that the disconnected components of the network are considered to be distinct sub-graphs. Therefore we have $d(A,B)=0$ for all pairs of disconnected components $A$ and $B$ by definition. Since we are only interested in disconnected networks we set the value of $d(A,B)=1$ for all $A$ and $B$. We also change the denominator to correctly normalize the measure for a disconnected network such that a connected network would have a measure zero and a network with no edges (a maximally dismantled network), the measure of dismantlement would be 1. $$max(\sum_{i<j}|V_i||V_j|) = \frac{n(n-1)}{2}$$  

Therefore for a graph $G=(V,E)$ with disconnected components given by $P = \{ V_0, V_1, .. ,V_k \}$, we obtain the following, $$q(P) = \frac{2}{n(n-1)}\sum_{i<j}|V_i||V_j| $$

\section{Results}
\label{S:2}
For a given partition $P= \{ V_0, V_1, .. ,V_k \}$ the amount of dismantlement must necessarily increase when the number of components is increased by further breaking down the connected components by removing the edges in the network. Suppose we obtain a partition $P^{'} = \{ V^{a}_0, V^{b}_0, V_1, .. ,V_k \}$ by breaking down $V_0$, then it can be easily shown that, $q(P)$ is strictly less than $q(P^{'})$.
$$ q(P^{'})-q(P) = \frac{2}{n(n-1)} |V^{a}_0||V^{b}_0| > 0$$
This notion cannot be incorporated if we use the size of the largest component as a measure of dismantlement, unless the largest component is being broken down by removing nodes or edges.  

Furthermore, for a graph of size $n$ and $k$ connected components, the amount of dismantlement must increase as the variance of the sizes of components must decreases, as components of uniform size would always be more dismantled when networks with equal number of components are compared. The amount of dismantlement in a disconnected network must increase as the number of disconnected components increases. The proposed measure incorporates this notion. For any two disconnected components the product of the sizes of these components is maximised when the sizes are equal to each other. The product decreases as the the difference between the sizes increases.

The measure of dismantlement of the networks in figure 1 is $0.5102$ for the first network and $0.7176$ for the second according to the proposed measure. 

\bibliography{main}
\bibliographystyle{unsrt}  
\end{document}